\documentstyle[preprint,aps,rotating]{revtex}
\draft
\begin{document}
\tightenlines
\input epsf
\title{\begin{flushright} {\rm\small HUB-EP-98/30}\end{flushright}
Exclusive semileptonic decays of $B$ mesons to orbitally excited $D$
mesons in the relativistic quark model}
\author{ D. Ebert$^1$, R. N. Faustov$^2$
~and V. O. Galkin$^1$\thanks{On leave of absence 
from Russian Academy of Sciences,
Scientific Council for Cybernetics,
Vavilov Street 40, Moscow 117333, Russia.}}

\address{$^1$Institut f\"ur Physik, Humboldt--Universit\"at zu Berlin,
Invalidenstr.110, D-10115 Berlin, Germany\\
$^2$Russian Academy of Sciences, Scientific Council for
Cybernetics, Vavilov Street 40, Moscow 117333, Russia}

\date{\today}
\maketitle
\begin{abstract}
Exclusive semileptonic $B$ meson decays to orbitally excited $D$ mesons
are investigated in the infinitely heavy quark limit in the framework
of the relativistic quark model based on the quasipotential approach.
The $B\to D^{**}e\nu$ Isgur-Wise functions $\tau_{3/2}(w)$
and $\tau_{1/2}(w)$ are determined. It is found
that the relativistic transformation of the meson wave functions (Wigner
rotation of the light quark spin) contribute already at the leading order
of the heavy quark expansion. 
\end{abstract}  

\bigskip

The investigation of semileptonic decays of $B$ mesons to excited
$D$ meson states is a problem which is important both from  a theoretical
and experimental point of view. In particular, these decays can 
provide an additional source of information for the determination of the 
Cabibbo-Kobayashi-Maskawa matrix element $V_{cb}$ as well as on the 
relativistic quark dynamics inside heavy-light meson. The experimental
data on these decays are becoming available now \cite{cleo,aleph,opal},
and the $B$ factories
will provide more accurate and comprehensive data. The presence of the heavy 
quark in the initial and final meson state in these decays considerably
simplifies their theoretical description. A good starting point in this
analysis is the infinitely heavy quark limit, $m_Q\to\infty$ \cite{iw}. 
In this limit
the heavy quark symmetry arises, which strongly
reduces the number of independent
weak decay form factors \cite{iw1}.
The heavy quark mass and spin decouple then and all meson properties are
determined by light degrees of freedom alone. As a result the heavy
quark degeneracy of levels emerges. The spin $s_q$ of 
the light quark couples with
its orbital momentum $l$ ($j=l\pm s_q$), resulting for $P$-wave mesons in 
two degenerate
$j=3/2$ states ($J^P=1^{+},2^+$)\footnote{ Here $J=j\pm 1/2$ and 
$P$ are the total
angular momentum and parity of the meson.} and two degenerate $j=1/2$ states 
($0^+,1^+$). The heavy quark symmetry also predicts that the weak decay
form factors for $B\to D^{**}e\nu$ decays, where $D^{**}$ is a generic 
$P$-wave $D$ meson state, can be expressed in terms of two independent
Isgur-Wise functions $\tau_{3/2}$ and $\tau_{1/2}$ \cite{iw1}.

In preceding papers we have calculated the mass spectra of 
orbitally and radially excited states of heavy-light mesons \cite{egf}
as well as different weak decays of $B$ mesons to ground state
heavy and light mesons \cite{fgm,efg} in the 
framework of the relativistic quark model based on the quasipotential
approach in quantum field theory. Let us now  apply this model
to the investigation of semileptonic $B$ decays to $P$-wave $D$ mesons
in the heavy quark limit.

In the quasipotential approach a meson is described by the wave
function of the bound quark-antiquark state, which satisfies the
quasipotential equation \cite{3} of the Schr\"odinger type~\cite{4}
\begin{equation}
\label{quas}
{\left(\frac{b^2(M)}{2\mu_{R}}-\frac{{\bf
p}^2}{2\mu_{R}}\right)\Psi_{M}({\bf p})} =\int\frac{d^3 q}{(2\pi)^3}
 V({\bf p,q};M)\Psi_{M}({\bf q}),
\end{equation}
where the relativistic reduced mass is
\begin{equation}
\mu_{R}=\frac{M^4-(m^2_q-m^2_Q)^2}{4M^3}.
\end{equation}
Here $m_{q,Q}$ are the masses of light
and heavy quarks, and ${\bf p}$ is their relative momentum.  
In the centre-of-mass system the relative momentum squared on mass shell 
reads
\begin{equation}
{b^2(M) }
=\frac{[M^2-(m_q+m_Q)^2][M^2-(m_q-m_Q)^2]}{4M^2}.
\end{equation}

The kernel 
$V({\bf p,q};M)$ in Eq.~(\ref{quas}) is the quasipotential operator of
the quark-antiquark interaction. It is constructed with the help of the
off-mass-shell scattering amplitude, projected onto the positive
energy states. An important role in this construction is played 
by the Lorentz-structure of the confining quark-antiquark interaction
in the meson.  In 
constructing the quasipotential of quark-antiquark interaction 
we have assumed that the effective
interaction is the sum of the usual one-gluon exchange term and the mixture
of vector and scalar linear confining potentials.
The quasipotential is then defined by
\cite{mass}  
\begin{eqnarray}
\label{qpot}
V({\bf p,q};M)&=&\bar{u}_q(p)
\bar{u}_Q(-p)\Bigg\{\frac{4}{3}\alpha_sD_{ \mu\nu}({\bf
k})\gamma_q^{\mu}\gamma_Q^{\nu}\cr
& & +V^V_{\rm conf}({\bf k})\Gamma_q^{\mu}
\Gamma_{Q;\mu}+V^S_{\rm conf}({\bf
k})\Bigg\}u_q(q)u_Q(-q),
\end{eqnarray}
where $\alpha_s$ is the QCD coupling constant, $D_{\mu\nu}$ is the
gluon propagator in the Coulomb gauge
and ${\bf k=p-q}$; $\gamma_{\mu}$ and $u(p)$ are 
the Dirac matrices and spinors
\begin{equation}
\label{spinor}
u^\lambda({p})=\sqrt{\frac{\epsilon(p)+m}{2\epsilon(p)}}
{1\choose \frac{\bbox{\sigma p}}{\epsilon(p)+m}}\chi^\lambda
\end{equation}
with $\epsilon(p)=\sqrt{{\bf p}^2+m^2}$.
The effective long-range vector vertex is
given by
\begin{equation}
\Gamma_{\mu}({\bf k})=\gamma_{\mu}+
\frac{i\kappa}{2m}\sigma_{\mu\nu}k^{\nu},
\end{equation}
where $\kappa$ is the Pauli interaction constant characterizing the
anomalous chromomagnetic moment of quarks. Vector and
scalar confining potentials in the nonrelativistic limit reduce to
\begin{eqnarray}
V^V_{\rm conf}(r)&=&(1-\varepsilon)Ar,\nonumber\\
V^S_{\rm conf}(r)& =&\varepsilon Ar+B,
\end{eqnarray}
reproducing 
\begin{equation}
V_{\rm conf}(r)=V^S_{\rm conf}(r)+
V^V_{\rm conf}(r)=Ar+B,
\end{equation}
where $\varepsilon$ is the mixing coefficient. 

The quasipotential for the heavy quarkonia,
expanded in $v^2/c^2$, can be found in Refs.~\cite{mass,pot} and for
heavy-light mesons in \cite{egf}.
All the parameters of
our model like quark masses, parameters of the linear confining potential,
mixing coefficient $\varepsilon$ and anomalous
chromomagnetic quark moment $\kappa$ were fixed from the analysis of
heavy quarkonia masses \cite{mass} and radiative decays \cite{gf}. 
The quark masses
$m_b=4.88$ GeV, $m_c=1.55$ GeV, $m_s=0.50$ GeV, $m_{u,d}=0.33$ GeV and
the parameters of the linear potential $A=0.18$ GeV$^2$ and $B=-0.30$ GeV
have usual quark model values.  The value of the vector-scalar mixing
coefficient $\varepsilon=-1$
has been determined from the consideration of the heavy quark expansion
\cite{fg} and meson radiative decays \cite{gf}.
Finally, the universal Pauli interaction constant $\kappa=-1$ has been
fixed from the analysis of the fine splitting of heavy quarkonia ${
}^3P_J$- states \cite{mass}. Note that the 
long-range  magnetic contribution to the potential in our model
is proportional to $(1+\kappa)$ and thus vanishes for the 
chosen value of $\kappa=-1$.

In order to calculate the exclusive semileptonic decay rate of the $B$ meson 
it is necessary to determine the corresponding matrix element of the 
weak current between meson states.
The matrix element of the weak current $J^W=\bar c\gamma_\mu(1-\gamma^5)b$ 
between $B$ meson and orbitally excited $D^{**}$ meson in
the quasipotential method has the form \cite{f}
\begin{equation}\label{mxet} 
\langle D^{**} \vert J^W_\mu (0) \vert B\rangle
=\int \frac{d^3p\, d^3q}{(2\pi )^6} \bar \Psi_{D^{**}}({\bf
p})\Gamma _\mu ({\bf p},{\bf q})\Psi_B({\bf q}),\end{equation}
where $\Gamma _\mu ({\bf p},{\bf
q})$ is the two-particle vertex function and  $\Psi_{B,D^{**}}$ are the
meson wave functions projected onto the positive energy states of
quarks and boosted to the moving reference frame.
 The contributions to $\Gamma$ come from Figs.~1 and 2.\footnote{  
The contribution $\Gamma^{(2)}$ is the consequence
of the projection onto the positive-energy states. Note that the form of the
relativistic corrections resulting from the vertex function
$\Gamma^{(2)}$ is explicitly dependent on the Lorentz-structure of the
$q\bar q$-interaction.} In the heavy quark limit
$m_{b,c}\to \infty$ only $\Gamma^{(1)}$ contributes, while $\Gamma^{(2)}$ 
contributes at $1/m_{Q}$ order. As we limit our analysis here to
the leading order
of the heavy quark expansion,
only the vertex function $\Gamma^{(1)}$ is necessary.
It looks like
\begin{equation} \label{gamma1}
\Gamma^{(1)}({\bf
p},{\bf q})=\bar u_{c}(p_c)\gamma_\mu(1-\gamma^5)u_b(q_b)
(2\pi)^3\delta({\bf p}_q-{\bf
q}_q),\end{equation}
where \cite{f} 
\begin{eqnarray*} 
p_{c,q}&=&\epsilon_{c,q}(p)\frac{p_{D^{**}}}{M_{D^{**}}}
\pm\sum_{i=1}^3 n^{(i)}(p_{D^{**}})p^i,\\
q_{b,q}&=&\epsilon_{b,q}(p)\frac{p_B}{M_B} \pm \sum_{i=1}^3 n^{(i)}
(p_B)q^i,\end{eqnarray*}
and
$$ n^{(i)\mu}(p)=\left\{ \frac{p^i}{M},\ \delta_{ij}+
\frac{p^ip^j}{M(E+M)}\right\}.$$

The wave function of a $P$-wave $D^{**}$ meson at rest is given by
\begin{equation}\label{psi}
\Psi_{D^{**}}({\bf p})\equiv
\Psi^{JM}_{D(j)}({\bf p})={\cal Y}^{JM}_j\psi_{D(j)}({\bf p}),
\end{equation}
where $J$ and $M$ are the total meson angular momentum and its projection,
while $j$ is the light quark angular momentum.   
$\psi_{D(j)}({\bf p})$ is the radial part of the wave function,
which has been determined by the numerical solution of eq.~(\ref{quas})
in \cite{egf}.
The spin-angular momentum part ${\cal Y}^{JM}_j$ has the following form
\begin{equation}\label{angl}
{\cal Y}^{JM}_j=\sum_{\sigma_Q\sigma_q}\langle j\, M-\sigma_Q,\  
\frac12\, \sigma_Q |J\, M\rangle\langle 1\, M-\sigma_Q-\sigma_q,\ 
\frac12\, \sigma_q |j\, M-\sigma_Q\rangle Y_{1}^{M-\sigma_Q-\sigma_q}
\chi_Q(\sigma_Q)\chi_q(\sigma_q).
\end{equation}
Here $\langle j_1\, m_1,\  j_2\, m_2|J\, M\rangle$ are Clebsch-Gordan 
coefficients, $Y_l^m$ are spherical harmonics, and $\chi(\sigma)$ (where 
$\sigma=\pm 1/2$) are spin wave functions,
$$ \chi\left(1/2\right)={1\choose 0}, \qquad 
\chi\left(-1/2\right)={0\choose 1}. $$

It is important to note that the wave functions entering the weak current
matrix element (\ref{mxet}) are not at rest in general. E.g., 
in the $B$ meson rest frame the $D^{**}$ meson is moving with the recoil
momentum ${\bf \Delta}$. The wave function
of the moving $D^{**}$ meson $\Psi_{D^{**}\,{\bf\Delta}}$ is connected 
with the $D^{**}$ wave function at rest $\Psi_{D^{**}\,{\bf 0}}\equiv
\Psi_{D(j)}$ by the transformation \cite{f}
\begin{equation}
\label{wig}
\Psi_{D^{**}\,{\bf\Delta}}({\bf
p})=D_c^{1/2}(R_{L{\bf\Delta}}^W)D_q^{1/2}(R_{L{
\bf\Delta}}^W)\Psi_{D^{**}\,{\bf 0}}({\bf p}),
\end{equation}
where $R^W$ is the Wigner rotation and   
the rotation matrix $D^{1/2}(R)$ in spinor representation is given by
\begin{equation}\label{d12}
{1 \ \ \,0\choose 0 \ \ \,1}D^{1/2}_{c,q}(R^W_{L{\bf\Delta}})=
S^{-1}({\bf p}_{c,q})S({\bf\Delta})S({\bf p}),
\end{equation}
where
$$
S({\bf p})=\sqrt{\frac{\epsilon(p)+m}{2m}}\left(1+\frac{\bbox{ \alpha p}}
{\epsilon(p)+m}\right)
$$
is the usual Lorentz transformation matrix of the four-spinor.
For electro-weak $B$ meson
decays to $S$-wave final mesons such transformation contributes at first
order of the $1/m_Q$ expansion, while for the decays to excited final mesons
it gives a contribution already to the leading term, due to the orthogonality
of the initial and final meson wave functions.
  
In the infinitely heavy quark limit ($m_{b,c}\to\infty$)
all form factors of the 
semileptonic $B\to D^{**}e\nu$ decays are related to two independent
Isgur-Wise form factors $\tau_{1/2}$ and $\tau_{3/2}$ by \cite{iw1}
\begin{eqnarray}
\label{formf}
\langle D_0(1/2)(v')|\bar c\gamma_\mu(1-\gamma^5)b|B(v)\rangle&=&2\sqrt{M_B
M_{D_0(1/2)}}\ \tau_{1/2}(w)(v'_\mu-v_\mu),\cr
\langle D_1(1/2)(v')|\bar c\gamma_\mu(1-\gamma^5)b|B(v)\rangle&=&2\sqrt{M_B
M_{D_1(1/2)}}\ \tau_{1/2}(w)\big\{i\varepsilon_{\mu\alpha\beta\gamma}
\epsilon^{*\alpha}v^\beta v'^\gamma\cr
&&+(1-w)\epsilon^*_\mu+(\epsilon^*\cdot v)v'_\mu\big\},\cr
\langle D_1(3/2)(v')|\bar c\gamma_\mu(1-\gamma^5)b|B(v)\rangle
&=&\sqrt{\frac{M_B
M_{D_1(3/2)}}{2}}\ \tau_{3/2}(w)\big\{i(w+1)\varepsilon_{\mu\alpha\beta\gamma}
\epsilon^{*\alpha}v^\beta v'^\gamma\cr
&&+(1-w^2)\epsilon^*_\mu-(\epsilon^*\cdot v)[3v_\mu-(w-2)v'_\mu]\big\},\cr
\langle D_2(3/2)(v')|\bar c\gamma_\mu(1-\gamma^5)b|B(v)\rangle&=&\sqrt{3M_B
M_{D_2(3/2)}}\ \tau_{3/2}(w)\big\{-i\varepsilon_{\mu\alpha\beta\gamma}
\epsilon^{*\alpha\eta}v_\eta v^\beta v'^\gamma\cr
&&+[(w+1)\epsilon^*_{\mu\alpha} v^\alpha-\epsilon^*_{\alpha\beta} 
v^\alpha v^\beta v'_\mu]\big\},
\end{eqnarray}
where $v(v')$ is the four-velocity of the initial (final) meson, 
$w=v\cdot v'$,
and $\epsilon^\mu$, $\epsilon^{\mu\nu}$ are polarization vector and tensor
of $D_1$ and $D_2$ mesons, respectively.

To calculate the corresponding matrix elements,
we substitute the vertex function $\Gamma^{(1)}$ (\ref{gamma1}) in the 
matrix element of the weak current between meson states (\ref{mxet}) and
take into account the wave function properties (\ref{psi})--(\ref{wig}).
Then, in the limit $m_{b,c}\to \infty$ we find that the heavy 
quark symmetry relations (\ref{formf}) are exactly satisfied in our model.
The resulting expressions for the Isgur-Wise functions $\tau_{3/2}$ and
$\tau_{1/2}$ are 
\begin{eqnarray}
\label{tau3}
\tau_{3/2}(w)&=&\frac{\sqrt{2}}{3}\frac{1}{(w+1)^{3/2}}\int\frac{d^3 p}
{(2\pi)^3}\bar\psi_{D(3/2)}({\bf p}+\frac{2\epsilon_q}{M_{D(3/2)}(w+1)}
{\bf \Delta})\cr
&&\times\left[-2\epsilon_q
\overleftarrow{\frac{\partial}{\partial p}}+\frac{p}{\epsilon_q+m_q}\right]
\psi_B({\bf p}),\\
\label{tau1}
\tau_{1/2}(w)&=&\frac{1}{3\sqrt{2}}\frac{1}{(w+1)^{1/2}}\int\frac{d^3 p}
{(2\pi)^3}\bar\psi_{D(1/2)}({\bf p}+\frac{2\epsilon_q}{M_{D(1/2)}(w+1)}
{\bf \Delta})\cr
&&\times\left[-2\epsilon_q
\overleftarrow{\frac{\partial}{\partial p}}-\frac{2p}{\epsilon_q+m_q}\right]
\psi_B({\bf p}),
\end{eqnarray}
where the arrow over $\partial/\partial p$ indicates that the derivative
acts on the wave function of the $D^{**}$ meson. The last  terms in 
the square brackets of these
expressions result from the wave function transformation (\ref{wig}) 
associated with the relativistic  rotation of the
light quark spin (Wigner rotation) in
passing to the moving reference frame. These terms are numerically important
and lead to the suppression of the $\tau_{1/2}$ form factor compared to 
$\tau_{3/2}$. Note that if we  
had applied a simplified nonrelativistic quark model \cite{iw1,vo}
these important contributions would be missing. Neglecting further the
small difference between the wave functions $\psi_{D(1/2)}$ and 
$\psi_{D(3/2)}$, the following relation between $\tau_{3/2}$ and 
$\tau_{1/2}$ would have been obtained \cite{llsw}
\begin{equation}\label{taunr}
\tau_{1/2}(w)=\frac{w+1}{2}\tau_{3/2}(w). 
\end{equation}
However, we see that this relation is violated if relativistic
transformation properties of wave function are taken into account. 
At the point $w=1$, where the initial $B$ meson and final $D^{**}$ are
at rest, we find instead the relation
\begin{equation}\label{diftau}
\tau_{3/2}(1)-\tau_{1/2}(1)\cong \frac12\int\frac{d^3p}{(2\pi)^3}
\bar\psi_{D^{**}}({\bf p})\frac{p}{\epsilon_q+m_q}\psi_B({\bf p}),
\end{equation}
obtained by assuming $\psi_{D(3/2)}\cong\psi_{D(1/2)}\cong\psi_{D^{**}}$.
This relation (\ref{diftau}) coincides with the one found in 
Ref.~\cite{mlopr} where the Wigner rotation was also taken into account. 

In Table~\ref{tauv} we present our numerical results for $\tau_{j}(1)$ 
and its slope $\left.\rho_j^2=-\frac{1}{\tau_j}\frac{\partial}{\partial w}
\tau_j\right|_{w=1}$ in comparison with other model predictions 
\cite{llsw,mlopr,ddgnp,w,cdp,gi,cccn}. Moreover, 
we plot $\tau_j(w)$ for $B\to D^{**}e\nu$ and $B_s\to D_s^{**}e\nu$ decays
as function of $w$
in Figs.~3,~4. The corresponding decay rates and branching
ratios are given in Tables~\ref{bd} and \ref{bsds}. We see that most of 
the above approaches predict close values for the function $\tau_{3/2}(1)$ 
and its slope $\rho_{3/2}^2$, while the results for $\tau_{1/2}(1)$ 
significantly differ from each other. This difference is a consequence
of a different treatment of the relativistic quark dynamics. Nonrelativistic
approaches predict $\tau_{3/2}(1)\simeq\tau_{1/2}(1)$ (see (\ref{taunr})),
while the relativistic treatment leads to $\tau_{3/2}(1)>\tau_{1/2}(1)$
(see (\ref{diftau})). Our results  for the branching ratios of 
$B\to D_{1,2}(3/2)e\nu$ decays are consistent with available experimental
data \cite{cleo,aleph}, which at present require to use some assumptions
about the branching fractions of the $D_J$ mesons.

Finally, let us test the fulfilment of the Bjorken 
sum rule \cite{b} in our model.
This sum rule states
\begin{equation}
\label{bsr}
\rho^2=\frac14+\sum_m|\tau^{(m)}_{1/2}(1)|^2 +2\sum_m |\tau^{(m)}_{3/2}(1)|^2
+\cdots ,
\end{equation}
where $\rho^2$ is the slope of the $B\to D^{(*)}e\nu$ Isgur-Wise function,
$\tau^{(m)}_{1/2}$
and $\tau^{(m)}_{3/2}$  are the form factors describing the
orbitally excited states discussed here and their
radial excitations, and  
ellipses denote contributions from non-resonant channels. 
We see that the contribution of the lowest lying $P$-wave
states ($m=0$) implies the bound 
\begin{equation}
\rho^2>\frac14 +|\tau_{1/2}(1)|^2+2|\tau_{3/2}(1)|^2=0.80,
\end{equation}  
which is in agreement with the slope $\rho^2=1.02$ in our model \cite{fg}.

In this paper we have applied the relativistic quark 
model to the consideration 
of semileptonic decays of $B$ mesons to orbitally 
excited charmed mesons in the
leading order of the heavy quark expansion. In particular,
it has been found that 
the Lorentz properties and transformations of meson wave functions play
an important role in the theoretical description of these decays. 
Thus, the 
Wigner rotation of the light quark spin gives a significant contribution 
already at the leading order of the heavy quark expansion.  In conclusion
let us mention that
the corrections in inverse powers of the
heavy quark mass $1/m_{c,b}$ to the decay rates
might turn out to be non-negligible, especially for spin
zero and spin one $D^{**}$ mesons \cite{llsw}. 
The investigation of such corrections
in the framework of our model is an important task that will be considered 
elsewhere.        

\smallskip
We thank  J.G. K\"orner and V.I. Savrin 
for useful discussions.
One of the authors (V.O.G) gratefully acknowledges the warm hospitality
of the colleagues in the particle theory group of the Humboldt-University
extended to him during his stay there.
He was supported in part by {\it Deutsche
Forschungsgemeinschaft} under contract Eb 139/1-3 and
in part by {\it Russian Foundation for Fundamental Research}
 under Grant No.\ 96-02-17171.
R.N.F. was supported in part by {\it Russian Foundation for
Fundamental Research} under Grant No.\ 96-02-17171.

\begin{table}
\caption{The comparison of our model results for
the values of the functions $\tau_j$ at zero 
recoil of final $D^{**}$ meson  and their slopes $\rho_j^2$ with other
predictions.} 
\label{tauv}
\begin{tabular}{cccccccc}
   & our & \cite{llsw} & \cite{ddgnp} & \cite{w} & \cite{cdp}& 
\cite{mlopr},\cite{gi} & \cite{mlopr},\cite{cccn}\\
\hline
$\tau_{3/2}(1)$ & 0.49 & 0.41 & 0.56 & 0.66 &     & 0.54 & 0.52\\
$\rho_{3/2}^2$  & 1.53 & 1.5  & 2.3  & 1.9  &     & 1.5  & 1.45\\
$\tau_{1/2}(1)$ & 0.28 & 0.41 & 0.09 & 0.41 &$0.35\pm0.08$ & 0.22 & 0.06\\
$\rho_{1/2}^2$  & 1.04 & 1.0  & 1.1  & 1.4  &$2.5\pm1.0$ & 0.83 & 0.73\\ 
\end{tabular}
\end{table}

\begin{table}
\caption{Decay rates $\Gamma$ (in units $|V_{cb}/0.04|^2 10^{-15}$ GeV) 
and branching ratios (\%) for $B\to D^{**}e\nu$ decays. }
\label{bd}
\begin{tabular}{ccccc}
Decay& $\Gamma$ & Br & Br (CLEO) \cite{cleo} & Br (ALEPH) \cite{aleph}\\
\hline
$B\to D_1(3/2)e\nu$ & 1.4 & 0.33& $0.56\pm 0.13\pm0.08\pm0.04$& $0.74\pm0.16$\\
$B\to D_2(3/2)e\nu$ & 2.1 & 0.52& $<0.8$ & $<0.2$\\
$B\to D_1(1/2)e\nu$ & 0.30 & 0.074 & &\\
$B\to D_0(1/2)e\nu$ & 0.25 & 0.062 & &\\
\end{tabular}
\end{table}

\begin{table}
\caption{Decay rates $\Gamma$ (in units $|V_{cb}/0.04|^2 10^{-15}$ GeV) 
and branching ratios (\%) for $B_s\to D_s^{**}e\nu$ decays.}
\label{bsds}
\begin{tabular}{ccc}
Decay& $\Gamma$ & Br \\
\hline
$B_s\to D_{s1}(3/2)e\nu$ & 1.6 & 0.39\\
$B_s\to D_{s2}(3/2)e\nu$ & 2.5 & 0.59\\
$B_s\to D_{s1}(1/2)e\nu$ & 0.54 & 0.13\\
$B_s\to D_{s0}(1/2)e\nu$ & 0.45 & 0.11 \\
\end{tabular}
\end{table}

\begin{figure}
\unitlength=0.9mm
\large
\begin{picture}(150,150)
\put(10,100){\line(1,0){50}}
\put(10,120){\line(1,0){50}}
\put(35,120){\circle*{8}}
\multiput(32.5,130)(0,-10){2}{\begin{picture}(5,10)
\put(2.5,10){\oval(5,5)[r]}
\put(2.5,5){\oval(5,5)[l]}\end{picture}}
\put(5,120){$b$}
\put(5,100){$\bar q$}
\put(5,110){$B$}
\put(65,120){$c$}
\put(65,100){$\bar q$}
\put(65,110){$D^{**}$}
\put(43,140){$W$}
\put(0,85){\small FIG. 1. Lowest order vertex function $\Gamma^{(1)}$
contributing to the current matrix element (9). }
\put(10,20){\line(1,0){50}}
\put(10,40){\line(1,0){50}}
\put(25,40){\circle*{8}}
\put(25,40){\thicklines \line(1,0){20}}
\multiput(25,40.5)(0,-0.1){10}{\thicklines \line(1,0){20}}
\put(25,39.5){\thicklines \line(1,0){20}}
\put(45,40){\circle*{2}}
\put(45,20){\circle*{2}}
\multiput(45,40)(0,-4){5}{\line(0,-1){2}}
\multiput(22.5,50)(0,-10){2}{\begin{picture}(5,10)
\put(2.5,10){\oval(5,5)[r]}
\put(2.5,5){\oval(5,5)[l]}\end{picture}}
\put(5,40){$b$}
\put(5,20){$\bar q$}
\put(5,30){$B$}
\put(65,40){$c$}
\put(65,20){$\bar q$}
\put(65,30){$D^{**}$}
\put(33,60){$W$}
\put(90,20){\line(1,0){50}}
\put(90,40){\line(1,0){50}}
\put(125,40){\circle*{8}}
\put(105,40){\thicklines \line(1,0){20}}
\multiput(105,40.5)(0,-0.1){10}{\thicklines \line(1,0){20}}
\put(105,39,5){\thicklines \line(1,0){20}}
\put(105,40){\circle*{2}}
\put(105,20){\circle*{2}}
\multiput(105,40)(0,-4){5}{\line(0,-1){2}}
\multiput(122.5,50)(0,-10){2}{\begin{picture}(5,10)
\put(2.5,10){\oval(5,5)[r]}
\put(2.5,5){\oval(5,5)[l]}\end{picture}}
\put(85,40){$b$}
\put(85,20){$\bar q$}
\put(85,30){$B$}
\put(145,40){$c$}
\put(145,20){$\bar q$}
\put(145,30){$D^{**}$}
\put(133,60){$W$}
\put(0,5){\small FIG. 2. Vertex function $\Gamma^{(2)}$
with the account of the quark
interaction. Dashed lines correspond  }
\put(0,0) {\small to the effective potential
(\ref{qpot}). Bold lines denote the negative-energy part of the quark
propagator. }

\end{picture}

\end{figure}
\setcounter{figure}2 

\begin{figure}
\centerline{\begin{turn}{-90}\epsfxsize=9.5cm 
\epsfbox{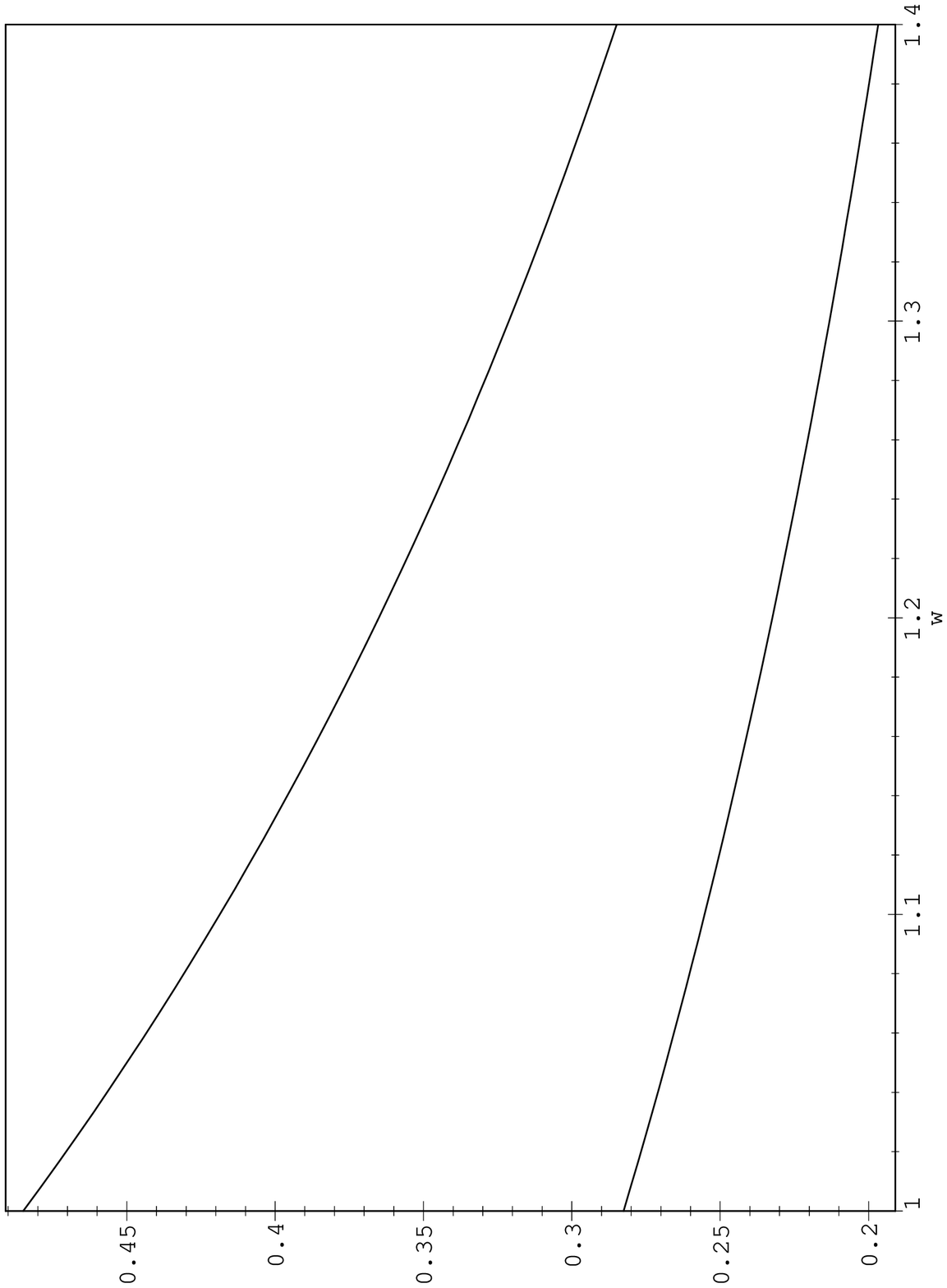}\end{turn}}
\caption{Isgur-Wise functions $\tau_{3/2}(w)$ (upper curve) and $\tau_{1/2}(w)$
(lower curve) for $B\to D^{**}e\nu$ decay.}
\label{ftaub}
\end{figure}

\begin{figure}
\centerline{\begin{turn}{-90}\epsfxsize=9.5cm 
\epsfbox{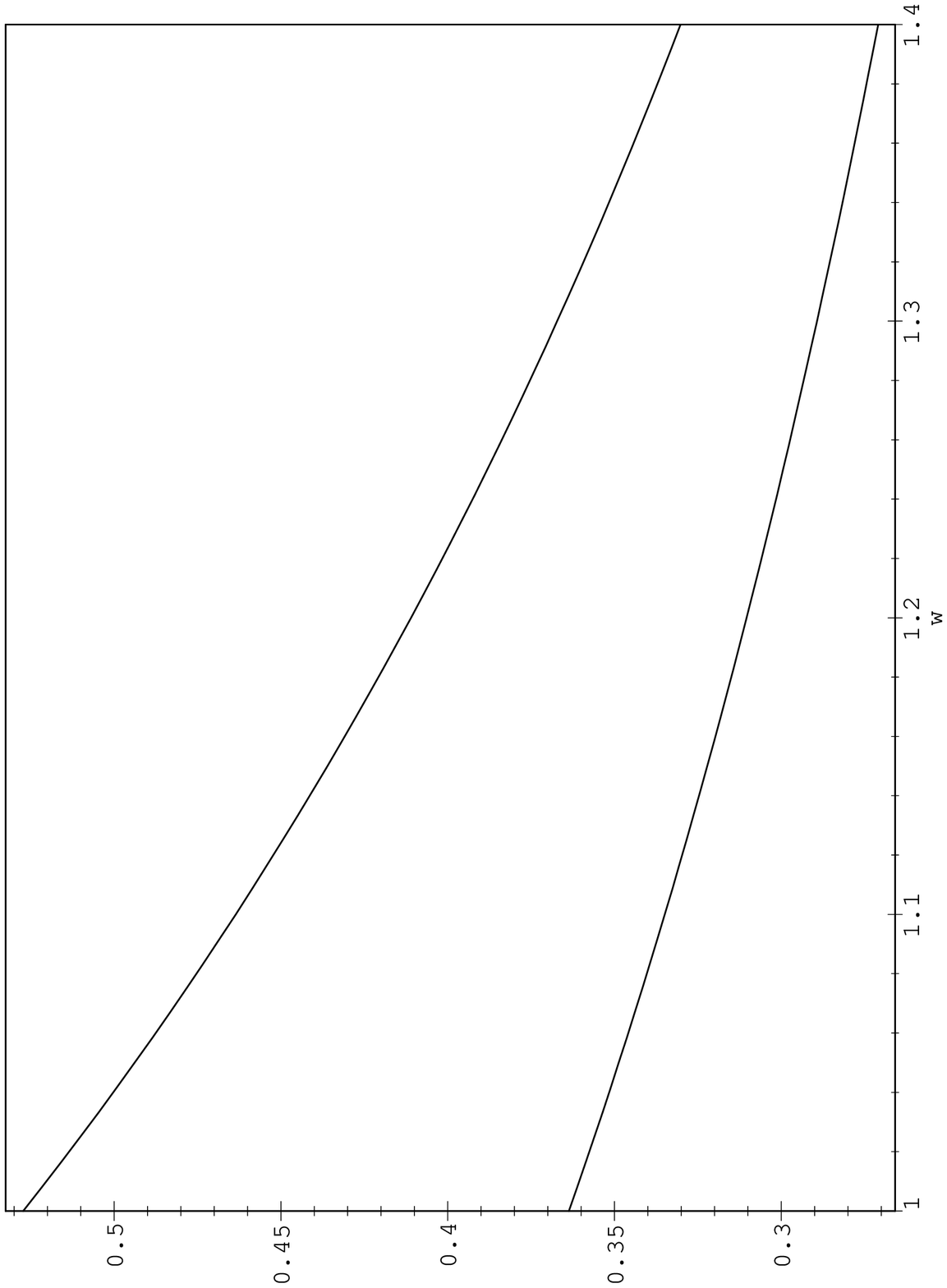}\end{turn}}
\caption{Isgur-Wise functions $\tau_{3/2}(w)$ (upper curve) and $\tau_{1/2}(w)$
(lower curve) for $B_s\to D_s^{**}e\nu$ decay.}
\label{ftaus}
\end{figure}

\end{document}